\documentclass[aps,prc,preprint,groupedaddress]{revtex4-1}

\usepackage{graphicx}
\usepackage{amssymb}

\begin{document}


\title{Pseudomagnetic effects for resonance neutrons in the search for time reversal invariance violation}



\author{Vladimir Gudkov}
\email[]{gudkov@sc.edu}
\affiliation{Department of Physics and Astronomy, University of South Carolina, Columbia, SC 29208}

\author{Hirohiko M. Shimizu}
\email[]{shimizu@phi.phys.nagoya-u.ac.jp}
\affiliation{Department of Physics, Nagoya University, Nagoya 464-8602, Japan}


\date{\today}

\begin{abstract}
A general theory of pseudomagnetic effects for  the propagation of polarized neutrons through polarized target in multi resonance approach is presented.
Some applications related to the proposed  search for time reversal invariance violation in neutron scattering are considered.
\end{abstract}


\maketitle


\section{Introduction}

Neutron spin optics in polarized nuclear targets became very important topic due to resent proposals to search for time reversal invariance violation (TRIV) in neutron nucleus scattering (see, for example \cite{Bowman:2014fca} and references therein).
These experiments require the understanding neuron spin dynamics while they propagate through polarized nuclear targets in the presence of a number of $s$-wave and $p$-wave resonances, since  neuron spin rotation due to strong spin-spin interactions can reduce the values of TRIV observable or, in some cases mimic TRIV effects \cite{BG:FSI,Gudkov:1992pm,Gudkov:1991qg,Masuda:1992,Lamoreaux:1994nd,Skoy:1996,Serebrov96,Masuda:1998zb,Lukashevich:2011zz}.

The phenomenon of neutron spin rotation, known as a pseudomagnetic effect, for a propagation of polarized slow neutrons through a polarized target was predicted in paper \cite{Baryshevsky:1964}. It is related to the fact that due to strong spin-spin interactions the value of the neutron wave index of refraction depends on the relative orientation of neutron spin in respect to the direction  of nuclear polarization
\begin{equation}\label{index}
n^2_{\pm}=1+\frac{4\pi}{k^2}\sum_iN_if^i_{\pm}.
\end{equation}
Here $N_i$ is the number of nuclei of type $i$ pur unit volume, $k$ is the neutron wave number, and $f^i_{\pm}$ is the neutron elastic forward scattering amplitude on an $i$-th nucleus for a positive and negative projection of the neutron spin on the direction of the nuclear polarization.
Then, taking into account that the second term in the above equation is much smaller than one, we can write the difference of the refractive indexes with different neutron spin orientation as
\begin{equation}\label{deltan}
\Delta n=n_+-n_-=\frac{2\pi}{k^2}\sum_iN_i(f^i_+-f^i_-) .
\end{equation}
This difference of refraction indices leads to the rotation of neutron spin around the direction of the nuclear polarization on an angle $\varphi =k\Delta n z$ while neutrons propagate through the target on the distance $z$.
Then, the corresponding frequency of the neutron spin precession can be written \cite{Baryshevsky:1964} as
\begin{equation}\label{PM}
\omega_{\rm P}=\frac{{\rm d} \varphi}{{\rm d}z}=\frac{2\pi N \hbar}{M_{\rm n}}\sum_iN_i{\rm Re}\, (f^i_+-f^i_-),
\end{equation}
where $M_{\rm n}$ is neutron mass.
For very low energies of  neutrons the scattering amplitudes do not depend on neutron energy, and as a consequence, the frequency $\omega_{\rm P}$ has a constant value which depends only on the properties of the polarized target.
Therefore, it was suggested \cite{Baryshevsky:1964} to use an effective pseudomagnetic field, which produces a precession of the neutron spin at the frequency $\omega_{\rm P}$,
\begin{equation}\label{HPM}
B_{\rm P}=\frac{\hbar \omega_{\rm P} }{2\mu_{\rm n}}
\end{equation}
as a natural  characteristic of the target (here $\mu_{\rm n}$ is the neutron magnetic moment). Numerically, $B_{\rm P}/(1{\rm T})=5.47 \omega_{\rm P}/(1 {\rm GHz})$.
These phenomenon was studied mostly for the case of very low energy (thermal) neutrons (see \cite{Abragam1973,abragam1982nuclear,Pokazanev1979,Tsulaia2014} and references therein).
In paper \cite{Forte:1973}, the pseudomagnetic spin precession has been studied in thepresence of a low energy $s$-wave neutron resonance.
In that case, the parameters $B_{\rm P}$ and $\omega_{\rm P}$ show very strong energy dependencies in the vicinity of the resonance.
In this paper we present a general formalism for the pseudomagnetic phenomena and apply it for the multi-resonance case of neutron resonances with different parities.

\section{General formalism for psedu-magnetic field}

Let us consider reaction matrix $\hat{T}$ which is related to the scattering matrix $\hat{\mathbb{S}}$ and the matrix $\hat{R}$ as
\begin{equation}\label{matr}
2\pi i \hat{T}=\hat{1}-\hat{\mathbb{S}}=\hat{R},
\end{equation}
thus the reaction amplitude $\hat{f}$ can be written as $\hat{f}=-\pi(k_{\rm i} k_{\rm f})^{-1/2}\hat{T}$, where $k_{\rm i,f}$ are values of initial and final momentum, correspondingly.
Then, to describe scattering of polarized neutrons on polarized target with spin $\vec{I}$, we need to calculate corresponding reaction matrix elements
\begin{equation}\label{Tmat}
\left\langle\vec{k}_{\rm f}\mu_{\rm f} \middle\vert T \middle\vert \vec{k}_{\rm i}\mu_{\rm i} \right\rangle ,
\end{equation}
where $\mu_{\rm i,f}$ is the projection of neutron spin on the axis of quantization.
For a coherent elastic scattering at zero angle the initial and final values of neutron momenta and spin projections are equal to each other, $\vec{k}_{\rm i}=\vec{k}_{\rm f}=\vec{k}$ and $\mu_{\rm i}=\mu_{\rm f}=\mu$.

It is convenient to relate this matrix to the matrix $\hat{R}$ in the integral of motion representation of $S$-matrix \cite{Baldin:1961}
\begin{equation}\label{Smat}
\left\langle S^{\prime} l^{\prime} \alpha^{\prime} \middle\vert \mathbb{S}^J \middle\vert S l \alpha \right\rangle
\delta_{JJ^{\prime}}\delta_{MM^{\prime}}\delta(E^{\prime} -E) ,
\end{equation}
where $J$ and $M$ are are total spin and its projection, $S$ is a spin channel, $l$ is an orbital momentum, and $\alpha$ corresponds to other internal quantum numbers.
Taking into account that the spin channel is a sum of neutron spin $\vec{s}$ nucleus spin $\vec{I}$
\begin{equation}\label{scheam1}
\vec{S}=\vec{s}+\vec{I},
\end{equation}
and the total spin is
\begin{equation}\label{schem2}
 \vec{J}=\vec{S}+\vec{l},
\end{equation}
one can write $T$-matrix elements as
\begin{eqnarray}\label{genMat}
 \nonumber
  2\pi i
  \left\langle \vec{k}\mu \middle\vert T \middle\vert \vec{k}\mu \right\rangle
  & & = \sum_{JMlml^{\prime} m^{\prime} S m_s S^{\prime} m^{\prime}_s}Y_{l^{\prime} m^{\prime}}(\theta ,\phi)
  	\left\langle s \mu I M_I \middle\vert S^{\prime} m^{\prime}_s \right\rangle
	\left\langle S^{\prime} m^{\prime}_s l^{\prime} m^{\prime} \middle\vert JM \right\rangle\\
  &\times&
  	\left\langle S^{\prime} l^{\prime} \alpha^{\prime} \middle\vert R^J \middle\vert S l \alpha\right\rangle
	\left\langle JM \middle\vert S m_s l m \right\rangle
	\left\langle S m_s \middle\vert s \mu I M_I \right\rangle
	Y^*_{lm} (\theta ,\phi) ,
 \end{eqnarray}
where angles $(\theta ,\phi)$ describe the direction of neutron momentum $\vec{k}$.
For simplicity of further formulae let's choose the quantization axis along the vector $\vec{k}$.
It should be noted, that for the case of $s$-wave neutrons all expressions do not depend on the choice of the quantization axis. For $p$-wave neutrons with an arbitrary choice of the quantization axis the formulas for the expressions became more complicated, but rather trivial ones.
Therefore, we  presented them in Appendix \ref{apxA}.
Then, the amplitude for neutron elastic scattering can be written as
 \begin{eqnarray}\label{ampl}
 \nonumber
f_{\mu } &=& \frac{i}{2k}\sum_{Jl  S  S^{\prime} M_I } (2l+1)
	\left\langle s \mu I M_I \middle\vert S^{\prime} m^{\prime}_s \right\rangle
	\left\langle S^{\prime} m^{\prime}_s l 0 \middle\vert J M \right\rangle \\
	&&\times
	\left\langle S^\prime l \middle\vert R^J \middle\vert S l \right\rangle
	\left\langle J M \middle\vert S m_s l 0 \right\rangle
	\left\langle S m_s \middle\vert s \mu I M_I \right\rangle .
 \end{eqnarray}
It should be noted that in the above expression for the amplitude the sum over $M_I$ must be taken carefully to satisfy the state of the polarization of the nuclear target.
For example, for the case of vector polarization, which we consider in details here, the only term with  $M_I=I$ is presented in the sum.

The matrix elements in Eq.(\ref{ampl}) for slow neutrons can be written in Breit-Wigner resonance approximation with one $s$-resonance or $p$-resonance as
\begin{eqnarray}\label{BWamp}
\left(F_{S'Sl}^{J}\right)_K
&\equiv&
\left\langle S^{\prime}_K l_K \middle\vert R^{J_K} \middle\vert S_K l_K \right\rangle
\nonumber\\
&=&
i \frac
{\sqrt{\Gamma^{\rm n}_{l_K}(S^{\prime}_K)}\sqrt{\Gamma^{\rm n}_{l_K}(S_K)}}
{E-E_K+i\Gamma_K/2}
e^{i(\delta_{l_K}(S^{\prime}_K)+\delta_{l_K}(S_K))}
-2ie^{i\delta_{l_K}(S_KS^{\prime}_K)}\sin\delta_{l_K}(S_K S^{\prime}_K)
\nonumber\\
\label{eq:1}
\end{eqnarray}
%
where $E_K$, $\Gamma_K$, and $\Gamma^n_{l_K}$ are the energy, the total width, and the partial neutron width of the $K$-th nuclear compound resonance, $E$ is the neutron energy, and $\delta_{l_K}$ is the potential scattering phase shift.
For $p$-wave resonances we keep only resonance term because for low energy neutrons $\delta_l\sim (kR_0)^{2l+1}$ (where $R_0$ is nucleus radius), and, as a consequence, a contribution from $p$-wave potential scattering is negligible.

Now, following the definition in  Eq.(\ref{PM}), one can obtain the frequency of neutron spin rotation due to pseudomagnetic field for a nuclear target with a single element  as
\begin{equation}\label{PM1}
\omega_{\rm P}=\frac{2\pi N \hbar}{M_{\rm n}}{\rm Re}\, (f_{\frac12}-f_{-\frac12}).
\end{equation}

One can see that for the case of $s$-wave neutron scattering on the vector polarized target, the difference of amplitudes in Eq.(\ref{PM1}) is
\begin{eqnarray}\label{ampDiff}
f_{\frac12}
-
f_{-\frac12}
=\frac{i}{2k}\frac{2I}{2I+1}\left(
F_{I+\frac12\,I+\frac12\,0}^{I+\frac12}
-
F_{I-\frac12\,I-\frac12\,0}^{I-\frac12}
\right).
\end{eqnarray}
For the case of very slow neutrons one can neglect the resonance term contribution (first term in Eq.(\ref{BWamp})) to $R$-matrices in the above equation.
Then $R$-matrix can be written in terms of neutron scattering lengths $a_{\pm}$ with spin orientations parallel and antiparallel to the direction of nuclear polarization as
\begin{equation}\label{ScLen}
\left\langle \left(I\pm \frac12\right) 0 \middle\vert R^{I\pm \frac12} \middle\vert \left(I\pm\frac12\right) 0 \right\rangle = -2 i k a_{\pm},
\end{equation}
which gives us the well known expression \cite{Baryshevsky:1964} for the pseudomagnetic frequency of thermal neutrons
\begin{equation}\label{PM0}
\omega_{\rm P}=\frac{4\pi N \hbar}{M_{\rm n}}\frac{I}{(2I+1)}(a_+-a_-).
\end{equation}
For low energy resonance region we need to take into account not only potential scattering, but also the contributions from each resonance.
Thus, for example, with the presence of  $s$-wave resonances with total spins $J=I\pm 1/2$, the
pseudomagnetic frequency became neutron energy dependent
\begin{eqnarray}
\label{PMs}
\omega^{\rm s}_{\rm P}
&=&
\frac{4\pi N \hbar}{M_{\rm n}}\frac{I}{(2I+1)}\left(
	a_+ - a_-
	-
	\sum_{K,\, l_K=0}
	\frac{\Gamma^{\rm n}_K}{2k}
	\frac
		{(E-E_K)}
		{(E-E_K)^2+(\Gamma_K/2)^2}
	\beta_K
\right),
\\
\beta_K &=& \left\{\begin{array}{ll}
	 1 & \quad (J_K=I+\frac12) \\
	-1 & \quad (J_K=I-\frac12) \\
	\end{array}\right.
\end{eqnarray}
where index $\pm$ for resonance parameters corresponds to resonances with total spins $J=I\pm 1/2$, correspondingly.
One can see that the pseudomagnetic frequency has a a sharp oscillation with sign changing at position of each $s$-wave resonance \cite{Forte:1973}.

For the case of $p$-wave resonances the corresponding difference of amplitudes in Eq.(\ref{PM1}) is
\begin{eqnarray}\label{PMpI}
&&f_{\frac12}
-
f_{-\frac12}=
\\
&=& \left\{ \begin{array}{ll}
	0
	& \quad (J=I-\frac32) \\
	-\frac{\displaystyle 3i}{\displaystyle k}\frac{\displaystyle I}{\displaystyle (2I+1)^2}\left(
		(2I-1) F_{I-\frac{1}{2}\,I-\frac{1}{2}\,1}^{J}
		+
		2 \frac{\displaystyle \sqrt{2I-1}}{\displaystyle \sqrt{I+1}} F_{I-\frac{1}{2}\,I+\frac{1}{2}\,1}^{J}
		+
		\frac{\displaystyle 1}{\displaystyle I+1} F_{I+\frac{1}{2}\,I+\frac{1}{2}\,1}^{J}
	\right)
	& \quad (J=I-\frac12) \\
	-\frac{\displaystyle 3i}{\displaystyle k}\frac{\displaystyle I}{\displaystyle (2I+1)^2}\left(
		2 F_{I-\frac{1}{2}\,I-\frac{1}{2}\,1}^{J}
		-
		2 \frac{2I-1}{\sqrt{I(2I+3)}} F_{I-\frac{1}{2}\,I+\frac{1}{2}\,1}^{J}
		-
		\frac{\displaystyle (5+4I)(I+1)}{\displaystyle 2I+3} F_{I+\frac{1}{2}\,I+\frac{1}{2}\,1}^{J}
	\right)
	& \quad (J=I+\frac12) \\
	\frac{\displaystyle 3i}{\displaystyle k} \frac{\displaystyle I}{\displaystyle (2I+3)(I+1)} F_{I+\frac{1}{2}\,I+\frac{1}{2}\,1}^{J}
	& \quad (J=I+\frac32) \nonumber \\
\end{array}\right.
\end{eqnarray}

 which leads to the pseudomagnetic frequency from $p$-resonances
\begin{eqnarray}
	\label{PMp}
\omega^{\rm p}_{\rm P} &=&
	\frac{6\pi N \hbar}{M_{\rm n}k}\frac{I}{(2I+1)}
	\sum_{K,\, l_K=1}
		\gamma_K
		\frac{E-E_K}
		{(E-E_K)^2+(\Gamma_K/2)^2}
	\\
\label{PMpgamma}
\gamma_K &=& \left\{\begin{array}{ll}
	0 & \quad (J_K=I-\frac32) \\
	\frac{\displaystyle 1}{\displaystyle 2I+1}
		\left(
		(2I-1)\sqrt{\Gamma^{\rm n}_K(I-\frac12)}\sqrt{\Gamma^{\rm n}_K(I-\frac12)}
		\right. & \\ \quad \left.
		+
		2\sqrt{\frac{\displaystyle 2I-1}{\displaystyle I+1}}
			\sqrt{\Gamma^{\rm n}_K(I+\frac12)}\sqrt{\Gamma^{\rm n}_K(I-\frac12)}
		\right. & \\ \quad \left.
		+
		2 \sqrt{\Gamma^{\rm n}_K(I+\frac12)}\sqrt{\Gamma^{\rm n}_K(I+\frac12)}
		\right)
		& \quad (J_K=I-\frac12) \\
	\frac{\displaystyle 1}{\displaystyle 2I+1}
		\left(
		2 \sqrt{\Gamma^{\rm n}_K(I-\frac12)}\sqrt{\Gamma^{\rm n}_K(I-\frac12)}
		\right. & \\ \quad \left.
		-
		2\sqrt{\frac{\displaystyle 2I-1}{\displaystyle I(2I+3)}}
			\sqrt{\Gamma^{\rm n}_K(I+\frac12)}\sqrt{\Gamma^{\rm n}_K(I-\frac12)}
		\right. & \\ \quad \left.
		-
		\frac{\displaystyle 5+4I(I+1)}{\displaystyle 2I+3}
			\sqrt{\Gamma^{\rm n}_K(I+\frac12)}\sqrt{\Gamma^{\rm n}_K(I+\frac12)}
		\right)
		& \quad (J_K=I+\frac12) \\
	- \frac{\displaystyle 2I+1}{\displaystyle (2I+3)(I+1)}
		\sqrt{\Gamma^{\rm n}_K(I+\frac12)}\sqrt{\Gamma^{\rm n}_K(I+\frac12)}
		& \quad (J_K=I+\frac32) \\
	\end{array}\right.
 \end{eqnarray}
  \nonumber
This expression looks complicated, however, since $p$-wave resonances are very weak in low energy region, usually only one most closed resonance contribution should be accounted.
Therefore, only three terms or less from the above expression will actually contribute to $p$-wave dependent part of the pseudomagnetic frequency.
Moreover, for the case of TRIV search only resonances with $J=I-1/2$ and $J=I+1/2$ present the interest, since only these resonances can be mixed with $s$-resonances (which have spins $J=I-1/2$ and $J=I+1/2$) by weak and TRIV interactions.
One can see also  that in contrast to $s$-wave resonances this pseudomagnetic frequency depends, in general, not on total neutron widths, but on the partial neutron widths for different spin channels.
(For relation of the spin channel formalism with the spin-orbital scheme formalism see Appendix \ref{apxC}.)

Up to now we considered the case of pure vector polarized mono isotopic target.
Based on the coherent nature of the pseudomagnetic effect, it is easy to generalized all above expressions for the case of composed target with an arbitrary polarization.
Thus, for composed (multi isotope) target, the total pseudomagnetic frequency is a linear sum of frequencies from all isotopes presented in the target.  The case of arbitrary polarization of each isotope is accounted by a summation of differences of amplitudes of Eq.(\ref{ampl}) taken with the corresponding weights $w(M_I)$ for each spin projection quantum number $M_I$, which is the weight in the density operator used for the description of general polarization in terms of the density polarization matrix.
Therefore, the resulting pseudomagnetic frequency $\omega^{*}_{P}$ can be written as \cite{Baryshevsky:1964}
\begin{equation}\label{polar}
\omega^{*}_{\rm P} = \omega_{\rm P} \frac{1}{I}\sum_{M_I} w(M_I)M_I.
\end{equation}

It should be noted that, in Eqs.(\ref{PMpI}) and (\ref{PMpgamma}), there are no contributions from the resonance with a total spin $J=I-3/2$, but there is a contribution with $J=I+3/2$.
This asymmetry simply reflects the fact that we consider the case with a pure vector polarization of the target, which corresponds to $M_I=I$.
For the case of mixed target polarization with a fractional population of the target nuclear level of $M_I=-I$, the resonance a spin $J=I-3/2$ can also lead to pseudomagnetic precession due to corresponding difference of amplitudes
\begin{equation}\label{resmin32}
f_{\frac12}-f_{-\frac12} = \frac{\displaystyle 3i}{\displaystyle k} \frac{\displaystyle I}{\displaystyle (2I+3)(I+1)} F_{I-\frac12\,I-\frac12\,1}^{J}.
\end{equation}
However, as it was mentioned above, these resonances cannot lead to TRIV effects.

\section{Pseudomagnetic effects in lanthanum aluminate}

Let us consider the application of the presented formalism for the pseudomagnetic effect of in Lanthanum Aluminate crystals. Since a very large parity violating effect was observed on $^{139}$La in the vicinity of $0.734$ eV resonance \cite{Alfimenkov:1982yv,Alfimenkov:1983sys,Shimizu:1992np,Mitchell2001157}, this isotope looks like a promising target for a search of TRIV effects in nuclei \cite{Bowman:2014fca}.
$^{139}$La nuclei can be polarized in lanthanum aluminate crystals with currently experimentally achieved value of $^{139}$La polarization \cite{Hautle:2000} of $47.5$\%.

\begin{figure}[h]
\includegraphics[width=86mm]{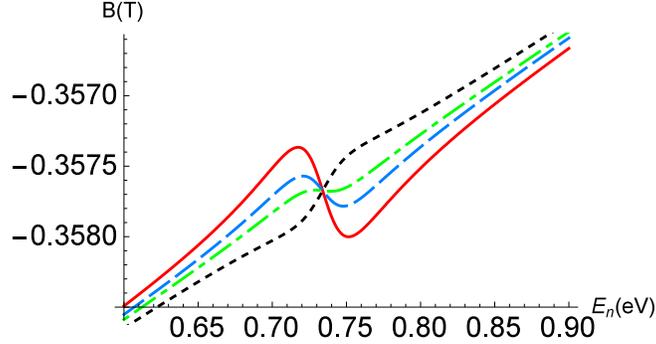}
\caption{ (Color online) Pseudomagnetic field in the fully polarized lanthanum target for $\alpha=0$ (solid line), $\alpha=\pi/4$ (dashed line), $\alpha=-\pi/4$ (dashed-dotted  line), and $\alpha=\pi/2$ (dotted line).}
\label{La}
\end{figure}

Since we do not know partial neutron widths for the $p$-wave resonance we describe the ratio $x_s= \sqrt{\Gamma^{\rm n}_{\rm p}(I-\frac12)}/\sqrt{\Gamma^{\rm n}_{\rm p}(I-\frac12)+\Gamma^{\rm n}_{\rm p}(I+\frac12)}$ (see Appendix \ref{apxC}) using a parameter $\alpha$, such that  $x_s=\sin\alpha$.
Fig.~\ref{La} shows the pseudomagnetic field in the 100\% polarized lanthanum target as a function of neutron energy in the vicinity of $p$-wave resonance for $\alpha=0$, $\alpha=\pi/4$, $\alpha=-\pi/4$, and $\alpha=\pi/2$.

\begin{figure}[h]
\includegraphics[width=86mm]{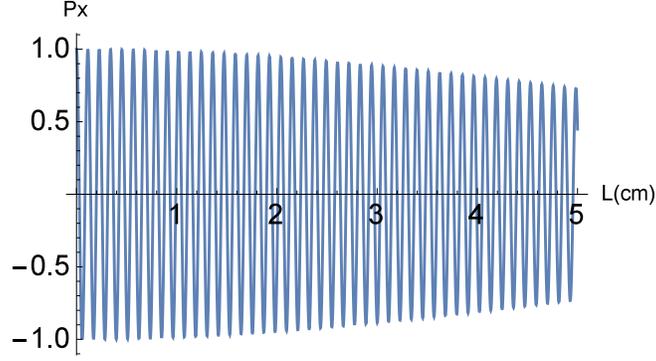}
\caption{ (Color online) Polarization of of $0.734$ eV neutrons in the fully polarized lanthanum  target as a function of the propagation distance $L$. }
\label{Polx}
\end{figure}

\begin{figure}[h]
\includegraphics[width=86mm]{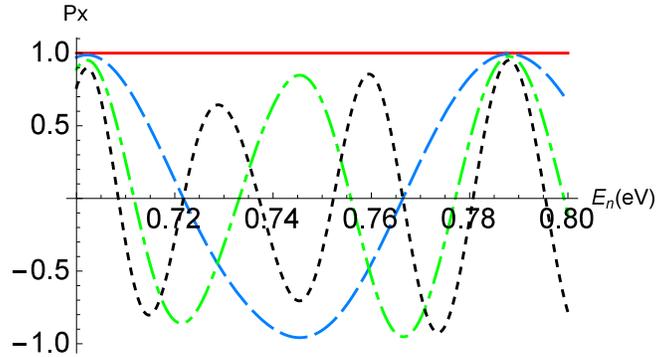}
\caption{ (Color online) Neutron polarization in the fully polarized lanthanum target as a function of neutron energy at $L=0$ cm (solid line), $L=2$ cm (dashed line), $L=4$ cm (dashed-dotted  line), and $L=6$ cm (dotted line).
}
\label{Poldist}
\end{figure}

Assuming that initial neutrons are polarized perpendicular to the quantization axis $z$ and along  the axis $x$, we can calculate neutron polarization $P_x$ (an expectation value of spin projection operator) as a function of the propagation distance $L$ in the target \cite{Baryshevsky:1964,Forte:1973}
\begin{equation}\label{polar}
P_x(L)=\frac
	{\cos\left(\frac{\displaystyle \omega_{\rm P}L}{\displaystyle v_{\rm n}}\right)}
	{{\rm cosh}\left(\frac{\displaystyle \omega' L}{\displaystyle v_{\rm n}}\right)},
\end{equation}
where $v_{\rm n}$ is neutron velocity and
\begin{equation}\label{kappa}
\omega' = \frac{2\pi N \hbar}{M_{\rm n}}{\rm Im}\, (f_{\frac12}-f_{-\frac12})
\end{equation}
is imaginary part of the pseudomagnetic frequency, which is related to neutron absorbtion in the target.
For the case of $La$ target with the parameter $\alpha=0$ and for neutron energy of $0.734$ eV, the polarization as a function of $L$ is shown in Fig. \ref{Polx}.
One can see that the value of neutron polarization is monotonously decreasing due to neutron absorbtion.
Fig.~\ref{Poldist} shows the polarization as a function of neutron energy at $L=0$ cm (solid line), $L=2$ cm (dashed line), $L=4$ cm (dashed-dotted  line), and $L=6$ cm (dotted line), which clearly demonstrate energy dependance of the pseudomagnetic effect.

\begin{figure}[h]
\includegraphics[width=86mm]{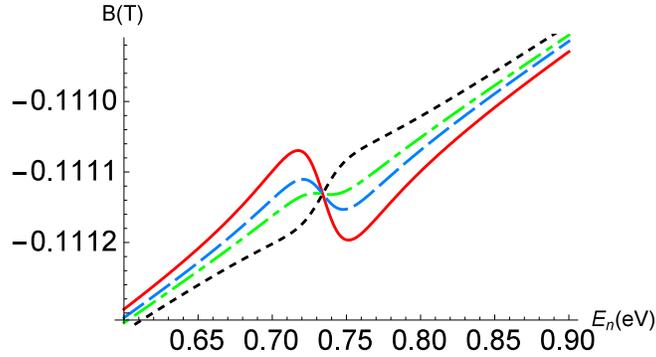}
\caption{ (Color online) Pseudomagnetic field in fully polarized LaAlO$_3$ target for $\alpha=0$ (solid line), $\alpha=\pi/4$ (dashed line), $\alpha=-\pi/4$ (dashed-dotted line), and $\alpha=\pi/2$ (dotted line).}
\label{LaCry}
\end{figure}

For the case of LaAlO$_3$ we need also include the pseudomagnetic field from polarized Al. Then, assuming a pure vector 100\% polarizations for both $^{139}$La and $^{27}$Al nuclei, the calculated  pseudomagnetic fields in the LaAlO$_3$ target for $\alpha=0$, $\alpha=\pi/4$, $\alpha=-\pi/4$, and $\alpha=\pi/2$  are shown in Fig.~\ref{LaCry}.

From the above pictures we can see that pseudomagnetic fields of La and Al in the vicinity of La $p$-wave resonance are oriented in opposite directions, moreover, the La pseudomagnetic field is opposite to the direction of the applied magnetic field used for the polarization of the target.  It demonstrates that, in principle, one can essentially reduce the pseudomagnetic field in a compound by choosing an appropriate combination of the elements with opposite directions of pseudomagnetic fields.

Here we discuss the case of LaAl$_x$X$_{1-x}$O$_3$ where Al is partially replaced by the element X.
Table~\ref{Tab1} shows the replacement fraction $x$ for X$=^{45}$Sc and X$=^{59}$Co to cancel the pseudomagnetic field of $^{139}$La.
It should be noted that the cancelation is calculated only at the thermal neutron energy neglecting all resonance contributions. In general, it strongly depends on the neutron energy.
As we can see from this table, for example, La and Al pseudomagnetic fields are parallel at the thermal neutron energy but have an opposite directions in the resonance region (see Figs. (\ref{LaCry}) and (\ref{La})).
It should be also noted that the additional absorption with the replacement of Al should be carefully considered in the design of the experiment.

\begin{table}
\caption{\label{Tab1} The replacement fraction $x$ for the cancelation of the pseudomagnetism in LaAl$_x$X$_{1-x}$O$_3$  at the thermal neutron energy.}
   \begin{tabular}{|c|c|c|c|c|}
	\hline
     Element & $I$ & Abundance & $(a_+-a_-) [{\rm fm}]$ & $x$ \\
      \hline
     $^{139}$La & 7/2 & 0.9991 & 6.9 & - \\
     $^{27}$Al & 5/2 & 1 & 0.52 & - \\
     $^{45}$Sc & 7/2 & 1 & -12.08 & 0.59 \\
     $^{59}$Co & 7/2 & 1 & -12.79 & 0.56 \\
    \hline
   \end{tabular}
\end{table}

\section{Conclusions}

The presented study of the pseudomagnetic spin rotation for the propagation of polarized neutrons through polarized targets shows the importance of multi resonance description of the effects.
We show that the effective pseudomagnetic field has a noticeable energy dependence in the vicinity of $p$-wave resonance and it is rather sensitive to target structure, to the polarization pattern of different nuclei in the target, and to the values of partial neutron widths.
Therefore, by changing composition materials of the target and by applying external magnetic field it is possible to reduce effect of the pseudomagnetic field in the given interval of the neutron energy for a particular target.
The partial neutron widths have been measured only using angular distribution measurements in neutron radiative capture.
The sensitivity of the pseudomagnetic field to the values of the partial neutron widths gives a new method to measure them in the neutron transmission through polarized targets.

\begin{acknowledgments}
This material is based upon work supported by the U.S. Department of Energy Office of Science, Office of Nuclear Physics program under Award Number DE-SC0015882.
\end{acknowledgments}

\appendix

\section{A general form for the difference of p-wave amplitudes}
\label{apxA}
The general formula for the difference of p-wave amplitudes amplitudes for a pure vector polarized target can be obtained fron Eq.(\ref{genMat}). Then, choosing direction of the target polarization along axis $z$ and momentum direction along vector $\vec{n}(\theta ,\phi )$ we obtain
\begin{eqnarray}\label{PMpGen}
&&f_{\frac12}
-
f_{-\frac12} =  \\
&=& \left\{ \begin{array}{ll}
	A^{-3/2}_{--}F_{I-\frac{1}{2}\,I-\frac{1}{2}\,1}^{J}
	& \quad (J=I-\frac32) \\
		A^{-1/2}_{--} F_{I-\frac{1}{2}\,I-\frac{1}{2}\,1}^{J}
		+
		A^{-1/2}_{-+}  F_{I-\frac{1}{2}\,I+\frac{1}{2}\,1}^{J}
		+
		A^{-1/2}_{++} F_{I+\frac{1}{2}\,I+\frac{1}{2}\,1}^{J}
	& \quad (J=I-\frac12) \\
		A^{1/2}_{--} F_{I-\frac{1}{2}\,I-\frac{1}{2}\,1}^{J}
		+
		A^{1/2}_{-+} F_{I-\frac{1}{2}\,I+\frac{1}{2}\,1}^{J}
		+
		A^{1/2}_{++} F_{I+\frac{1}{2}\,I+\frac{1}{2}\,1}^{J}
	& \quad (J=I+\frac12) \\
	A^{3/2}_{++} F_{I+\frac{1}{2}\,I+\frac{1}{2}\,1}^{J}
	& \quad (J=I+\frac32) \nonumber \\
\end{array}\right.
\end{eqnarray}

Where spin-angular coefficients are given by
\begin{equation}\label{ammm3}
A^{-3/2}_{--}
	=
	-
	\frac{3i}{2k}
	\frac{I-1}{2I+1}
	\sin^2\theta
\end{equation} ,
for $I>1$,
\begin{equation}\label{ammm1}
A^{-1/2}_{--}
	=
	-
	\frac{3i}{k}
	\frac
	{I\left(\sin^2\theta+\sqrt{2I-1}\sin2\theta\cos\phi+(2I-1)\cos^2\theta\right)}
	{(2I+1)^2} ,
\end{equation}

\begin{equation}\label{Amsum1}
A^{-1/2}_{-+}
	=
	\frac{3i}{k}
	\frac
	{I\left(
		(2I-3)\sin\theta\cos\theta\cos\phi
		+\sqrt{2I-1}\left(
			1-3\cos^2\theta
		\right)
	\right)}
	{\sqrt{I+1}(2I+1)^2} ,
\end{equation}

\begin{equation}\label{Apmp1}
A^{-1/2}_{++}
	=
	\frac{3i}{2k}
	\frac{I}{(I+1)(2I+1)^2}
	\left(\sqrt{2 I-1} \sin2\theta\cos\phi+(I+1)(2I-1) \sin^2\theta-2\cos2\theta \right),
\end{equation}

\begin{eqnarray}\label{Ampm1} \nonumber
A^{1/2}_{--}
	&=&
	-
	\frac{3i}{2k}
	\frac{1}{(2I+1)^2} \left[
		4I\cos^2\theta
		+ 2I\sqrt{2I+1}\sin2\theta\cos\phi
		\right. \\
		&+& \left.
		\left(2I^2+I+1-2\sqrt{2 I+1}\sqrt{I}\cos2\phi\right)\sin^2\theta
		+
		2\sqrt{I}\sin2\theta\cos\phi
	\right] ,
\end{eqnarray}

\begin{eqnarray}\label{ApSum1} \nonumber
A^{1/2}_{-+}
	&=&
	-\frac{3i}{2k}
	\frac{1}{(2I+1)^2\sqrt{2I+3}}
  	\left[
		-4 \sqrt{I}(2I-1)\cos^2\theta
		\right.\\
		&+&2 (2I-1) \left(
			\sqrt{2I+1}\cos2\phi+\sqrt{I}
		\right)\sin^2\theta \\
		&+& \left.
		\left\{ \sqrt{I}\sqrt{2I+1}(2I-3)-(6I-1) \right\} \sin2\theta\cos\phi
	\right] , \nonumber
\end{eqnarray}

\begin{eqnarray}\label{Appp1} \nonumber
A^{1/2}_{++}
	&=&
	\frac{3i}{k}
	\frac{1}{(2I+1)^2(2I+3)}
	\left[
		I(4I^2+4I+5) \cos^2\theta\right. \\ \nonumber
   &+&  \left(\sqrt{2I+1}(2I^2+I+1)-(2I-1)\sqrt{I}\right)\sin2\theta\cos\phi \\
   &+& \left.  \left(I(2I-1)-2\sqrt{I}\sqrt{2I+1}\cos2\phi\right) \sin^2\theta
   \right] ,
\end{eqnarray}

\begin{eqnarray}\label{Appp3}
A^{3/2}_{++}
	&=&
	\frac{3i}{2k} \frac{1}{(I+1)(2I+3)} \left[
  		2I\cos^2\theta
		\right.\nonumber\\
		&+&
		\left\{
			\frac{I\left(2I^2+5I+5\right)}{2I+1}
	   		+
			\left(
				\sqrt{\frac{3(I+1)}{(2I+1)}}
				-
				\sqrt{I+1}\sqrt{2I+3}
			\right)
			\cos2\phi
		\right\} \sin^2\theta
		\nonumber\\
		&+&\left.
		\left(
			2\sqrt{I+1}
			-\sqrt{2I+3}(I+1)
			-\sqrt{\frac{3}{2I+1}}
		\right) \sin2\theta\cos\phi
	\right] .
	\nonumber\\
\end{eqnarray}

\section{Relations between different spin-coupling schemes}
\label{apxC}

The relation between two spin coupling schemes $\vec{J}=(\vec{I}+\vec{s})+\vec{l}$ and $\vec{J}=(\vec{s}+\vec{l})+\vec{I}$ is given by
\begin{equation}\label{spincoupl}
\left\langle
	\left(\left(l,\frac12\right)j,I\right)J
\middle\vert
	\left(l,\left(\frac12,I\right)S\right)J
\right\rangle
	=(-1)^{l+I+j+S}\sqrt{(2j+1)(2S+1)}\left\{
		\begin{array}{ccc}
			l & \frac12 & j \\
			I & J & S
		\end{array}
	\right\},
\end{equation}
where $\vec{S}=\vec{I}+\vec{s}$ and $\vec{j} =\vec{s}+\vec{l}$.

Now, defining
\begin{eqnarray}
  x &=& \left\vert j=\frac12 \right\rangle \\
  \nonumber
  y &=& \left\vert j=\frac32 \right\rangle \\
  \nonumber
  x_s &=& \left\vert S=I-\frac12 \right\rangle \\
  \nonumber
  y_s &=& \left\vert S=I+\frac12 \right\rangle ,
\end{eqnarray}
one can write for $l=1$
\begin{eqnarray}
x_s &=& (-1)^{2I+1}\sqrt{4I}\left\{
	\begin{array}{ccc}
		1 & \frac{1}{2}& \frac{1}{2} \\
		I & J & I-\frac12
	\end{array}
	\right\} x
	+(-1)^{2I}\sqrt{8I}\left\{
	\begin{array}{ccc}
		1 & \frac{1}{2} & \frac{3}{2} \\
		I & J & I-\frac12
	\end{array}
	\right\}y\\
 	\nonumber
y_s &=& (-1)^{2I}\sqrt{4(I+1)}\left\{
	\begin{array}{ccc}
		1 & \frac{1}{2} & \frac{1}{2} \\
		I & J & I+\frac12
	\end{array}
	\right\} x
	+(-1)^{2I+1}\sqrt{8(I+1)}\left\{
	\begin{array}{ccc}
		1 & \frac{1}{2} & \frac{3}{2} \\
		I & J & I+\frac12
	\end{array}
	\right\}y .
\end{eqnarray}

\section{Spin-operator representation}
\label{apxD}

Sometime for description of neutron propagation through a polarized target it is convenient  to use a spin operator  \cite{Gurevich,Landau:3}
 \begin{equation}\label{spOper}
\hat{f}=a+b(\vec{s}\cdot \vec{I})
\end{equation}
whose eigenvalues for $J=I\pm 1/2$ are   scattering amplitudes $f_{\pm 1/2}$. In that case, one can calculate  the coefficients $a$ and $b$ as
\begin{eqnarray}\label{ab}
 \nonumber
  a &=& \frac{1}{2I+1} \left[ (I+1)f_{\frac12}+If_{-\frac12}\right] \\
  b &=& \frac{2}{2I+1} (f_{\frac12}-f_{-\frac12}),
\end{eqnarray}
with amplitudes $f_{\pm\frac12}$ given in Eq.(\ref{ampl}).


\bibliography{TViolation,ParityViolation}

\begin{thebibliography}{24}%
\makeatletter
\providecommand \@ifxundefined [1]{%
 \@ifx{#1\undefined}
}%
\providecommand \@ifnum [1]{%
 \ifnum #1\expandafter \@firstoftwo
 \else \expandafter \@secondoftwo
 \fi
}%
\providecommand \@ifx [1]{%
 \ifx #1\expandafter \@firstoftwo
 \else \expandafter \@secondoftwo
 \fi
}%
\providecommand \natexlab [1]{#1}%
\providecommand \enquote  [1]{``#1''}%
\providecommand \bibnamefont  [1]{#1}%
\providecommand \bibfnamefont [1]{#1}%
\providecommand \citenamefont [1]{#1}%
\providecommand \href@noop [0]{\@secondoftwo}%
\providecommand \href [0]{\begingroup \@sanitize@url \@href}%
\providecommand \@href[1]{\@@startlink{#1}\@@href}%
\providecommand \@@href[1]{\endgroup#1\@@endlink}%
\providecommand \@sanitize@url [0]{\catcode `\\12\catcode `\$12\catcode
  `\&12\catcode `\#12\catcode `\^12\catcode `\_12\catcode `\%12\relax}%
\providecommand \@@startlink[1]{}%
\providecommand \@@endlink[0]{}%
\providecommand \url  [0]{\begingroup\@sanitize@url \@url }%
\providecommand \@url [1]{\endgroup\@href {#1}{\urlprefix }}%
\providecommand \urlprefix  [0]{URL }%
\providecommand \Eprint [0]{\href }%
\providecommand \doibase [0]{http://dx.doi.org/}%
\providecommand \selectlanguage [0]{\@gobble}%
\providecommand \bibinfo  [0]{\@secondoftwo}%
\providecommand \bibfield  [0]{\@secondoftwo}%
\providecommand \translation [1]{[#1]}%
\providecommand \BibitemOpen [0]{}%
\providecommand \bibitemStop [0]{}%
\providecommand \bibitemNoStop [0]{.\EOS\space}%
\providecommand \EOS [0]{\spacefactor3000\relax}%
\providecommand \BibitemShut  [1]{\csname bibitem#1\endcsname}%
\let\auto@bib@innerbib\@empty
\bibitem [{\citenamefont {Bowman}\ and\ \citenamefont
  {Gudkov}(2014)}]{Bowman:2014fca}%
  \BibitemOpen
  \bibfield  {author} {\bibinfo {author} {\bibfnamefont {J.~D.}\ \bibnamefont
  {Bowman}}\ and\ \bibinfo {author} {\bibfnamefont {V.}~\bibnamefont
  {Gudkov}},\ }\href {\doibase 10.1103/PhysRevC.90.065503} {\bibfield
  {journal} {\bibinfo  {journal} {Phys. Rev.}\ }\textbf {\bibinfo {volume}
  {C90}},\ \bibinfo {pages} {065503} (\bibinfo {year} {2014})}\BibitemShut
  {NoStop}%
\bibitem [{\citenamefont {Bunakov}\ and\ \citenamefont
  {Gudkov}(1984)}]{BG:FSI}%
  \BibitemOpen
  \bibfield  {author} {\bibinfo {author} {\bibfnamefont {V.~E.}\ \bibnamefont
  {Bunakov}}\ and\ \bibinfo {author} {\bibfnamefont {V.~P.}\ \bibnamefont
  {Gudkov}},\ }\href@noop {} {\bibfield  {journal} {\bibinfo  {journal} {J.
  Phys.(Paris) Colloq.}\ }\textbf {\bibinfo {volume} {45}},\ \bibinfo {pages}
  {C3} (\bibinfo {year} {1984})}\BibitemShut {NoStop}%
\bibitem [{\citenamefont {Gudkov}(1992{\natexlab{a}})}]{Gudkov:1992pm}%
  \BibitemOpen
  \bibfield  {author} {\bibinfo {author} {\bibfnamefont {V.~P.}\ \bibnamefont
  {Gudkov}},\ }\href@noop {} {\bibfield  {journal} {\bibinfo  {journal}
  {Phys.Rev.}\ }\textbf {\bibinfo {volume} {C46}},\ \bibinfo {pages} {357}
  (\bibinfo {year} {1992}{\natexlab{a}})}\BibitemShut {NoStop}%
\bibitem [{\citenamefont {Gudkov}(1992{\natexlab{b}})}]{Gudkov:1991qg}%
  \BibitemOpen
  \bibfield  {author} {\bibinfo {author} {\bibfnamefont {V.~P.}\ \bibnamefont
  {Gudkov}},\ }\href@noop {} {\bibfield  {journal} {\bibinfo  {journal} {Phys.
  Rept.}\ }\textbf {\bibinfo {volume} {212}},\ \bibinfo {pages} {77} (\bibinfo
  {year} {1992}{\natexlab{b}})}\BibitemShut {NoStop}%
\bibitem [{\citenamefont {Masuda}\ and\ \citenamefont
  {et~al.}(1992)}]{Masuda:1992}%
  \BibitemOpen
  \bibfield  {author} {\bibinfo {author} {\bibfnamefont {Y.}~\bibnamefont
  {Masuda}}\ and\ \bibinfo {author} {\bibnamefont {et~al.}},\ }in\ \href@noop
  {} {\emph {\bibinfo {booktitle} {Proceedings of WEIN '92}}},\ \bibinfo
  {editor} {edited by\ \bibinfo {editor} {\bibfnamefont {T.~D.}\ \bibnamefont
  {Vylov}}}\ (\bibinfo  {publisher} {World Scientific, Singapore},\ \bibinfo
  {year} {1992})\BibitemShut {NoStop}%
\bibitem [{\citenamefont {Lamoreaux}\ and\ \citenamefont
  {Golub}(1994)}]{Lamoreaux:1994nd}%
  \BibitemOpen
  \bibfield  {author} {\bibinfo {author} {\bibfnamefont {S.}~\bibnamefont
  {Lamoreaux}}\ and\ \bibinfo {author} {\bibfnamefont {R.}~\bibnamefont
  {Golub}},\ }\href@noop {} {\bibfield  {journal} {\bibinfo  {journal}
  {Phys.Rev.}\ }\textbf {\bibinfo {volume} {D50}},\ \bibinfo {pages} {5632}
  (\bibinfo {year} {1994})}\BibitemShut {NoStop}%
\bibitem [{\citenamefont {Skoy}(1996)}]{Skoy:1996}%
  \BibitemOpen
  \bibfield  {author} {\bibinfo {author} {\bibfnamefont {V.~R.}\ \bibnamefont
  {Skoy}},\ }\href@noop {} {\bibfield  {journal} {\bibinfo  {journal} {Phys.
  Rev. D}\ }\textbf {\bibinfo {volume} {53}},\ \bibinfo {pages} {4070}
  (\bibinfo {year} {1996})}\BibitemShut {NoStop}%
\bibitem [{\citenamefont {Serebrov}(1996)}]{Serebrov96}%
  \BibitemOpen
  \bibfield  {author} {\bibinfo {author} {\bibfnamefont {A.~P.}\ \bibnamefont
  {Serebrov}},\ }in\ \href@noop {} {\emph {\bibinfo {booktitle} {"Parity and
  Time Reversal Violation in Compound Nuclear States"}}},\ \bibinfo {editor}
  {edited by\ \bibinfo {editor} {\bibfnamefont {N.}~\bibnamefont {Auerbach}}\
  and\ \bibinfo {editor} {\bibfnamefont {J.~D.}\ \bibnamefont {Bowman}}}\
  (\bibinfo  {publisher} {World Scientific},\ \bibinfo {year} {1996})\ pp.\
  \bibinfo {pages} {327--333}\BibitemShut {NoStop}%
\bibitem [{\citenamefont {Masuda}(1998)}]{Masuda:1998zb}%
  \BibitemOpen
  \bibfield  {author} {\bibinfo {author} {\bibfnamefont {Y.}~\bibnamefont
  {Masuda}},\ }\href@noop {} {\bibfield  {journal} {\bibinfo  {journal}
  {Nucl.Phys.}\ }\textbf {\bibinfo {volume} {A629}},\ \bibinfo {pages} {479C}
  (\bibinfo {year} {1998})}\BibitemShut {NoStop}%
\bibitem [{\citenamefont {Lukashevich}\ \emph {et~al.}(2011)\citenamefont
  {Lukashevich}, \citenamefont {Aldushchenkov},\ and\ \citenamefont
  {Dallman}}]{Lukashevich:2011zz}%
  \BibitemOpen
  \bibfield  {author} {\bibinfo {author} {\bibfnamefont {V.}~\bibnamefont
  {Lukashevich}}, \bibinfo {author} {\bibfnamefont {A.}~\bibnamefont
  {Aldushchenkov}}, \ and\ \bibinfo {author} {\bibfnamefont {D.}~\bibnamefont
  {Dallman}},\ }\href@noop {} {\bibfield  {journal} {\bibinfo  {journal}
  {Phys.Rev.}\ }\textbf {\bibinfo {volume} {C83}},\ \bibinfo {pages} {035501}
  (\bibinfo {year} {2011})}\BibitemShut {NoStop}%
\bibitem [{\citenamefont {Baryshevsky}\ and\ \citenamefont
  {Podgoretsky}(1964)}]{Baryshevsky:1964}%
  \BibitemOpen
  \bibfield  {author} {\bibinfo {author} {\bibfnamefont {V.}~\bibnamefont
  {Baryshevsky}}\ and\ \bibinfo {author} {\bibfnamefont {M.}~\bibnamefont
  {Podgoretsky}},\ }\href@noop {} {\bibfield  {journal} {\bibinfo  {journal}
  {Zh.Eksp.Teor.Fiz.}\ }\textbf {\bibinfo {volume} {47}},\ \bibinfo {pages}
  {1050} (\bibinfo {year} {1964})},\ \bibinfo {note} {[Sov. Phys, JETP 20, 704
  (1965)]}\BibitemShut {NoStop}%
\bibitem [{\citenamefont {Abragam}\ \emph {et~al.}(1973)\citenamefont
  {Abragam}, \citenamefont {Bacchella}, \citenamefont {Gl\"atti}, \citenamefont
  {Meriel}, \citenamefont {Pinot},\ and\ \citenamefont
  {Piesvaux}}]{Abragam1973}%
  \BibitemOpen
  \bibfield  {author} {\bibinfo {author} {\bibfnamefont {A.}~\bibnamefont
  {Abragam}}, \bibinfo {author} {\bibfnamefont {G.~L.}\ \bibnamefont
  {Bacchella}}, \bibinfo {author} {\bibfnamefont {H.}~\bibnamefont {Gl\"atti}},
  \bibinfo {author} {\bibfnamefont {P.}~\bibnamefont {Meriel}}, \bibinfo
  {author} {\bibfnamefont {M.}~\bibnamefont {Pinot}}, \ and\ \bibinfo {author}
  {\bibfnamefont {J.}~\bibnamefont {Piesvaux}},\ }\href {\doibase
  10.1103/PhysRevLett.31.776} {\bibfield  {journal} {\bibinfo  {journal} {Phys.
  Rev. Lett.}\ }\textbf {\bibinfo {volume} {31}},\ \bibinfo {pages} {776}
  (\bibinfo {year} {1973})}\BibitemShut {NoStop}%
\bibitem [{\citenamefont {Abragam}\ and\ \citenamefont
  {Goldman}(1982)}]{abragam1982nuclear}%
  \BibitemOpen
  \bibfield  {author} {\bibinfo {author} {\bibfnamefont {A.}~\bibnamefont
  {Abragam}}\ and\ \bibinfo {author} {\bibfnamefont {M.}~\bibnamefont
  {Goldman}},\ }\href@noop {} {\emph {\bibinfo {title} {Nuclear magnetism:
  order and disorder}}},\ International series of monographs on physics\
  (\bibinfo  {publisher} {Clarendon Press},\ \bibinfo {year}
  {1982})\BibitemShut {NoStop}%
\bibitem [{\citenamefont {Pokazan'ev}\ and\ \citenamefont
  {Skrotskii}(1979)}]{Pokazanev1979}%
  \BibitemOpen
  \bibfield  {author} {\bibinfo {author} {\bibfnamefont {V.~G.}\ \bibnamefont
  {Pokazan'ev}}\ and\ \bibinfo {author} {\bibfnamefont {G.~V.}\ \bibnamefont
  {Skrotskii}},\ }\href {\doibase 10.1070/PU1979v022n12ABEH005653} {\bibfield
  {journal} {\bibinfo  {journal} {Physics-Uspekhi}\ }\textbf {\bibinfo {volume}
  {22}},\ \bibinfo {pages} {943} (\bibinfo {year} {1979})}\BibitemShut
  {NoStop}%
\bibitem [{\citenamefont {Tsulaia}(2014)}]{Tsulaia2014}%
  \BibitemOpen
  \bibfield  {author} {\bibinfo {author} {\bibfnamefont {M.~I.}\ \bibnamefont
  {Tsulaia}},\ }\href {\doibase 10.1134/S1063778814100123} {\bibfield
  {journal} {\bibinfo  {journal} {Physics of Atomic Nuclei}\ }\textbf {\bibinfo
  {volume} {77}},\ \bibinfo {pages} {1321} (\bibinfo {year}
  {2014})}\BibitemShut {NoStop}%
\bibitem [{\citenamefont {Forte}(1973)}]{Forte:1973}%
  \BibitemOpen
  \bibfield  {author} {\bibinfo {author} {\bibfnamefont {M.}~\bibnamefont
  {Forte}},\ }\href@noop {} {\bibfield  {journal} {\bibinfo  {journal} {Il
  Nuov. Cim.}\ }\textbf {\bibinfo {volume} {18A}},\ \bibinfo {pages} {726}
  (\bibinfo {year} {1973})}\BibitemShut {NoStop}%
\bibitem [{\citenamefont {Baldin}\ \emph {et~al.}(1961)\citenamefont {Baldin},
  \citenamefont {Goldanskii},\ and\ \citenamefont {Rozental}}]{Baldin:1961}%
  \BibitemOpen
  \bibfield  {author} {\bibinfo {author} {\bibfnamefont {A.~M.}\ \bibnamefont
  {Baldin}}, \bibinfo {author} {\bibfnamefont {V.~I.}\ \bibnamefont
  {Goldanskii}}, \ and\ \bibinfo {author} {\bibfnamefont {I.~L.}\ \bibnamefont
  {Rozental}},\ }\href@noop {} {\emph {\bibinfo {title} {Kinematics of Nuclear
  Reactions}}}\ (\bibinfo  {publisher} {Oxford University Press},\ \bibinfo
  {address} {New York},\ \bibinfo {year} {1961})\BibitemShut {NoStop}%
\bibitem [{\citenamefont {Alfimenkov}\ \emph {et~al.}(1982)\citenamefont
  {Alfimenkov}, \citenamefont {Borzakov}, \citenamefont {Vo}, \citenamefont
  {Mareev}, \citenamefont {Pikelner}, \citenamefont {Khrykin},\ and\
  \citenamefont {Sharapov}}]{Alfimenkov:1982yv}%
  \BibitemOpen
  \bibfield  {author} {\bibinfo {author} {\bibfnamefont {V.~P.}\ \bibnamefont
  {Alfimenkov}}, \bibinfo {author} {\bibfnamefont {S.~B.}\ \bibnamefont
  {Borzakov}}, \bibinfo {author} {\bibfnamefont {V.~T.}\ \bibnamefont {Vo}},
  \bibinfo {author} {\bibfnamefont {{\relax Yu}.~D.}\ \bibnamefont {Mareev}},
  \bibinfo {author} {\bibfnamefont {L.~B.}\ \bibnamefont {Pikelner}}, \bibinfo
  {author} {\bibfnamefont {A.~S.}\ \bibnamefont {Khrykin}}, \ and\ \bibinfo
  {author} {\bibfnamefont {E.~I.}\ \bibnamefont {Sharapov}},\ }\href@noop {}
  {\bibfield  {journal} {\bibinfo  {journal} {JETP Lett.}\ }\textbf {\bibinfo
  {volume} {35}},\ \bibinfo {pages} {51} (\bibinfo {year} {1982})},\ \bibinfo
  {note} {[Pisma Zh. Eksp. Teor. Fiz.35,42(1982)]}\BibitemShut {NoStop}%
\bibitem [{\citenamefont {Alfimenkov}\ \emph {et~al.}(1983)\citenamefont
  {Alfimenkov}, \citenamefont {Borzakov}, \citenamefont {Van~Thuan},
  \citenamefont {Mareev}, \citenamefont {Pikelner}, \citenamefont {Khrykin},\
  and\ \citenamefont {Sharapov}}]{Alfimenkov:1983sys}%
  \BibitemOpen
  \bibfield  {author} {\bibinfo {author} {\bibfnamefont {V.~P.}\ \bibnamefont
  {Alfimenkov}}, \bibinfo {author} {\bibfnamefont {S.~B.}\ \bibnamefont
  {Borzakov}}, \bibinfo {author} {\bibfnamefont {V.}~\bibnamefont {Van~Thuan}},
  \bibinfo {author} {\bibfnamefont {{\relax Yu}.~D.}\ \bibnamefont {Mareev}},
  \bibinfo {author} {\bibfnamefont {L.~B.}\ \bibnamefont {Pikelner}}, \bibinfo
  {author} {\bibfnamefont {A.~S.}\ \bibnamefont {Khrykin}}, \ and\ \bibinfo
  {author} {\bibfnamefont {E.~I.}\ \bibnamefont {Sharapov}},\ }\href {\doibase
  10.1016/0375-9474(83)90649-8} {\bibfield  {journal} {\bibinfo  {journal}
  {Nucl. Phys.}\ }\textbf {\bibinfo {volume} {A398}},\ \bibinfo {pages} {93}
  (\bibinfo {year} {1983})}\BibitemShut {NoStop}%
\bibitem [{\citenamefont {Shimizu}\ \emph {et~al.}(1993)\citenamefont
  {Shimizu}, \citenamefont {Adachi}, \citenamefont {Ishimoto}, \citenamefont
  {Masaike}, \citenamefont {Masuda},\ and\ \citenamefont
  {Morimoto}}]{Shimizu:1992np}%
  \BibitemOpen
  \bibfield  {author} {\bibinfo {author} {\bibfnamefont {H.~M.}\ \bibnamefont
  {Shimizu}}, \bibinfo {author} {\bibfnamefont {T.}~\bibnamefont {Adachi}},
  \bibinfo {author} {\bibfnamefont {S.}~\bibnamefont {Ishimoto}}, \bibinfo
  {author} {\bibfnamefont {A.}~\bibnamefont {Masaike}}, \bibinfo {author}
  {\bibfnamefont {Y.}~\bibnamefont {Masuda}}, \ and\ \bibinfo {author}
  {\bibfnamefont {K.}~\bibnamefont {Morimoto}},\ }\href {\doibase
  10.1016/0375-9474(93)90494-I} {\bibfield  {journal} {\bibinfo  {journal}
  {Nucl. Phys.}\ }\textbf {\bibinfo {volume} {A552}},\ \bibinfo {pages} {293}
  (\bibinfo {year} {1993})}\BibitemShut {NoStop}%
\bibitem [{\citenamefont {Mitchell}\ \emph {et~al.}(2001)\citenamefont
  {Mitchell}, \citenamefont {Bowman}, \citenamefont {Penttilä},\ and\
  \citenamefont {Sharapov}}]{Mitchell2001157}%
  \BibitemOpen
  \bibfield  {author} {\bibinfo {author} {\bibfnamefont {G.}~\bibnamefont
  {Mitchell}}, \bibinfo {author} {\bibfnamefont {J.}~\bibnamefont {Bowman}},
  \bibinfo {author} {\bibfnamefont {S.}~\bibnamefont {Penttilä}}, \ and\
  \bibinfo {author} {\bibfnamefont {E.}~\bibnamefont {Sharapov}},\ }\href
  {\doibase http://dx.doi.org/10.1016/S0370-1573(01)00016-3} {\bibfield
  {journal} {\bibinfo  {journal} {Physics Reports}\ }\textbf {\bibinfo {volume}
  {354}},\ \bibinfo {pages} {157 } (\bibinfo {year} {2001})}\BibitemShut
  {NoStop}%
\bibitem [{\citenamefont {Hautle}\ and\ \citenamefont
  {Iinuma}(2000)}]{Hautle:2000}%
  \BibitemOpen
  \bibfield  {author} {\bibinfo {author} {\bibfnamefont {P.}~\bibnamefont
  {Hautle}}\ and\ \bibinfo {author} {\bibfnamefont {M.}~\bibnamefont
  {Iinuma}},\ }\href@noop {} {\bibfield  {journal} {\bibinfo  {journal}
  {Nucl.Instrum.Meth.}\ }\textbf {\bibinfo {volume} {A440}},\ \bibinfo {pages}
  {638} (\bibinfo {year} {2000})}\BibitemShut {NoStop}%
\bibitem [{\citenamefont {Gurevich}\ and\ \citenamefont
  {Tarasov}(1968)}]{Gurevich}%
  \BibitemOpen
  \bibfield  {author} {\bibinfo {author} {\bibfnamefont {I.~I.}\ \bibnamefont
  {Gurevich}}\ and\ \bibinfo {author} {\bibfnamefont {L.~V.}\ \bibnamefont
  {Tarasov}},\ }\href@noop {} {\emph {\bibinfo {title} {Low-energy neutron
  physics}}}\ (\bibinfo  {publisher} {Amsterdam, North-Holland Pub. Co.},\
  \bibinfo {year} {1968})\BibitemShut {NoStop}%
\bibitem [{\citenamefont {Landau}\ and\ \citenamefont
  {Lifshitz}(1981)}]{Landau:3}%
  \BibitemOpen
  \bibfield  {author} {\bibinfo {author} {\bibfnamefont {L.~D.}\ \bibnamefont
  {Landau}}\ and\ \bibinfo {author} {\bibfnamefont {E.}~\bibnamefont
  {Lifshitz}},\ }\href@noop {} {\emph {\bibinfo {title} {Quantum Mechanics
  (Non-relativistic Theory), third ed.}}}\ (\bibinfo  {publisher}
  {Butterworth-Heinemann, London},\ \bibinfo {year} {1981})\BibitemShut
  {NoStop}%
\end{thebibliography}%

\end{document}